\documentclass[a4paper,12pt]{article}

\usepackage{amssymb,amsmath,mathbbol,mathrsfs}
\usepackage[usenames,dvipsnames]{color}
\usepackage{hyperref}
\usepackage{xcolor}
\usepackage{stmaryrd}
\usepackage{authblk}
\usepackage{framed}
\usepackage{empheq}
\usepackage{slashed}
\usepackage{hyphenat}
\usepackage{cite}
\usepackage{apacite}
\usepackage{natbib}
\usepackage[normalem]{ulem} 
\usepackage{soul}

\usepackage{dsfont}
\usepackage{chngcntr}
\usepackage{enumerate}

\usepackage{marginnote}    

\usepackage[left=.8in,right=.8in,top=.8in,bottom=.8in]{geometry}                  

\usepackage{sectsty}
\allsectionsfont{\sffamily\mdseries\upshape} 
\usepackage{tocloft}

\makeatletter
\renewenvironment{abstract}{%
    \if@twocolumn
      \section*{\abstractname}%
    \else 
      \begin{center}%
        {\bfseries\sffamily\abstractname\vspace{\z@}}
      \end{center}%
      \quotation
    \fi}
    {\if@twocolumn\else\endquotation\fi}
\makeatother

\numberwithin{equation}{section}
5

\hypersetup{
	colorlinks=true,         
	linkcolor=MidnightBlue,          
	citecolor=BrickRed,        
	urlcolor=MidnightBlue            
}

\newcommand\mathC{\mkern1mu\raise2.2pt\hbox{$\scriptscriptstyle|$}
        {\mkern-7mu\rm C}}              

\newcommand{\be}{\begin{equation}}
\newcommand{\ee}{\end{equation}}

\renewcommand{\d}{{\mathrm{d}}}

\newcommand{\cint}{{\int\kern-.87em{<}}}
\newcommand{\sint}{{\int\kern-.75em{\sim}}}
\newcommand{\fint}{{\int\kern-1.00em{\int}}}

\let\oldmarginpar\marginpar
\renewcommand\marginpar[1]{\oldmarginpar{\color{red}\raggedright\footnotesize #1}}

\begin{document}

\title{The Hole Argument and Beyond:\\
 Part I: The Story so Far}
\author{Henrique Gomes and Jeremy Butterfield\footnote{\href{mailto:gomes.ha@gmail.com}{gomes.ha@gmail.com, jb56@cam.ac.uk}} \\\it Trinity College, University of Cambridge; Oriel College, University of Oxford }

\maketitle
\vspace{-1cm}
\begin{abstract}
  In this two-part paper, we review, and then develop, the assessment of the hole argument for general relativity. This first Part reviews the literature hitherto, focussing on the philosophical aspects. It also introduces two main ideas we will need in Part II: which will propose a framework for making comparisons of {\em non}-isomorphic spacetimes.

In Section 1 of this paper, we recall Einstein’s original argument. Section 2 recalls the argument's revival by philosophers in the 1980s and 1990s. This includes the first main idea we will need in Part II: namely, that two spacetime points in different possible situations are never strictly {\em identical}—they are merely {\em counterparts}. 

 In Section 3, we report---and rebut---more recent claims to ``dissolve” the argument. Our rebuttal is based on the fact that in differential geometry, and its applications in physics such as general relativity, points are in some cases identified, or correspond with each other, between one context and another, by means {\em other than} isometry (or isomorphism). We call such a correspondence a {\em threading} of points. This is the second main idea we shall use in Part II.

 \end{abstract}
  
\newpage

\tableofcontents 

\newpage

 \section{Act 1: Introduction}\label{intro}
 
 \subsection{Our two main aims}\label{ssec2aims}
 This two-part paper has two main aims. Our first aim (here in Part I) is to give a brisk review of the philosophical discussion of the hole argument for general relativity during the last thirty-five years. This discussion centres around whether and how one spacetime point in a model (solution) of general relativity corresponds to (or: is identical to; or should be identified with) a spacetime point in another such model that is isomorphic to the given one. This isomorphism is given by a diffeomorphism on the first model’s base-manifold of points. For in the hole argument, the second model, whose existence poses the threat of indeterminism, is defined by dragging along the first model’s metric and other fields: the field-values get ``re-distributed” on the manifold’s points.\footnote{So the first model can lack any automorphisms, i.e. isomorphisms to itself: it can be ``wrinkly’’. Cf. Section \ref{ssecEin13}.}
 
The second paper (Part II) addresses our second, more positive, aim. This is about the question of how one should make comparisons of {\em non}-isomorphic models, i.e. solutions of general relativity: in particular, how to do so, based on comparing points between the two models.  We will propose a framework for such comparisons. It proceeds by analogy with field-space formulations of gauge theories. It  postulates a fibre bundle whose fibres are isomorphic copies of a given model, i.e. a given Lorentzian manifold endowed with matter fields obeying the Einstein equations. So given a model based on a manifold $M$, Diff($M$) is the structure group of the bundle: any element of Diff($M$) drags the given model into coincidence (and so isomorphism) with another element of the fibre.  

This framework will build on Part I’s review, and especially on two main ideas that we develop here in Part I. The first (in Section \ref{sssec:cpart}) is David Lewis’ {\em counterpart theory} framework for treating the identity of objects---any objects, not only spacetime points---in different possible situations (in philosophical jargon: different {\em possible worlds}). Lewis says: objects in different possible worlds are never {\em identical}. Rather, they are {\em counterparts} of each other---better or worse counterparts, according to how similar they are. Although this seems puzzling at first sight, Lewis defends it convincingly---and we apply it to spacetime points: and in Part II, to points in non-isomorphic spacetimes.
 
 The second idea (in Section \ref{subsec:32thread}) is what we will call {\em threading} points: (it occurs between isomorphic spacetimes, and between non-isomorphic ones). For we will argue, {\em pace} some recent authors, that one cannot dismiss the hole argument as a confusion (or similarly: as a non-starter, or a {\em canard}), by claiming that isomorphism is the mandatory standard for when points in (the manifolds of) two models of the theory correspond (or: are identical, or should be identified with each other). For, we claim, isomorphism is {\em not} such a mandatory standard. Various examples show that both mathematics and physics treat the correspondence or identity of points flexibly---not always identifying them by isomorphism. We will call such a flexible correspondence or ``identification” a {\em threading}.

This paper is organised by invoking a dramatic metaphor: as three Acts, in three Sections.  Thus Act 1, in the rest of this Section, will review Einstein’s original version of the argument:   his version is an argument against general covariance (!). Act 2 (i.e. Section 2) recalls the revival of the argument by philosophers  in the 1980s and 1990s. This began with  \cite{EarmanNorton1987}'s casting the argument as against the philosophical doctrine {\em `substantivalism'}.  Like all `isms' in philosophy, `substantivalism' is a bit vague in its meaning; but the broad idea is that spacetime is, or is represented by, a manifold, and exists independently of its material content,  including the metric. In Section 2, we will take substantivalism to be the claim that spacetime points are objects; and we will recall some replies to Earman and Norton that were defences of substantivalism in this sense: including replies that invoked the idea of counterparts. Then Act 3 (i.e. Section 3) will report---and rebut---claims made since 2016 (first by \cite{Weatherall_hole}, and then by others)
that facts about mathematical practice rule out the hole argument. As we mentioned above, the idea of Weatherall et al. is that isomorphism is the mandatory standard for judging whether two points in (the manifolds of) two models of the theory correspond, or ``are identical”. Our rebuttal is based on the fact that mathematics and physics treat identity more flexibly: in differential geometry, and its applications such as general relativity, points are in some cases identified, or correspond with each other, between one context and another, by means {\em other than} isometry (or isomorphism): by what we will call a {\em threading}.

Two final orienting remarks. (1): Although the hole argument
focusses on 
whether two isomorphic models of general relativity  are, or represent,  the same physical possibility (``possible world”): the recent philosophical literature discusses this claim more generally, i.e. for an arbitrary physical theory that is formulated precisely enough to have a notion of isomorphism for its models, and for which we   interpret
such models as representing possibilities. Indeed,  this more general claim is widely endorsed (and so will be prominent in Acts 2 and 3), and  there is a  popular name for it. Namely, it is common to call the claim that any two isomorphic models of a theory are, or represent, the same physical possibility: {\em Sophistication}. So Acts 2 and 3 are in effect a review of the prospects for `Sophistication’. 

(2): Besides, this review of Sophistication connects with our second, positive, aim. For in other work, one of us \cite{Samediff_0, Samediff_1a, Samediff_1b} defends Sophistication for general relativity and for classical gauge theories (i.e. Yang-Mills theory, abelian or non-abelian), using  parallel reasoning in these two endeavours. So this other work treats Sophistication independently of the hole argument that we focus on here. And since all agree that it is natural to use fibre bundles in
 gauge theories, the fact that this other work treats general relativity and gauge theories in parallel makes it natural to use fibre bundles to discuss
 the ``trans-world identification” of points in general relativity.   Indeed, Part II's fibre bundle of models of general relativity will be a close cousin of field-space in gauge theories. In short: the parallels between general relativity and gauge theories, developed in the other work, support what we will do in Part II.

 \subsection{Einstein 1913}\label{ssecEin13}  
Einstein’s original hole argument, from the summer of 1913, was aimed {\em against} general covariance. At first sight, this target is of course very surprising, given Einstein’s desire in those years to achieve a generally covariant theory of gravitation, and his eventual success in doing so, in November 1915. But the explanation is now well-known; as follows. Einstein and Grossmann had just published their {\em Entwurf} theory. It had many features of general relativity, that Einstein wanted, and that in late 1915 he successfully implemented in his field equations.
 But the equations of the {\em Entwurf} theory are not generally covariant.\footnote{\label{history}{The stumbling block, in short (and in modern jargon), was that Einstein and Grossmann believed that a static spacetime must be spatially flat. The modern understanding of the history is mainly due to work in the 1980s by Stachel and Norton. Cf. e.g. \cite[ p. 267-284]{Norton1984}, \cite{Stachel_hole}, and \cite[Sec. 3.1]{Stachel_lrr}
}.} 

Pondering this defect, Einstein---with his characteristic, and brilliant, flexibility---thought: `maybe what I cannot get, I should anyway not want … maybe there is a good argument against general covariance’. And thus he thought of the hole argument. 

As Einstein conceives the argument, it urges that general covariance implies a radical indeterminism---which is unacceptable, so that general covariance must be rejected. The idea of the argument is that  if a theory of the type he sought---in which spacetime is not a ``fixed canvas”, or ``background arena”, on which material events play out, but is instead dynamically affected by such events---were to be generally covariant, then the theory  would be radically indeterministic: whatever the details of its field equations. For the state throughout spacetime except for a small patch, e.g. except for some small bounded open spacetime region---called `the hole’, though it is not topologically a hole---would not determine the state in that patch. The argument proceeds by using general covariance to construct two solutions of the putative theory (again: regardless of its detailed equations), with the following features: (i) they match exactly, for both metric and material fields, throughout spacetime outside the hole; but, according to the argument, (ii) they differ about the state of affairs inside the hole. 

We will present the argument briefly, and in modern terms. For it will be clear in later Sections that the differences from the historical Einstein will not matter for either of our two aims.\footnote{\label{history2}{From the vast literature on the history, we recommend, in addition to items in the previous footnote: \cite{Norton_EinsteinCov}, and \cite{Norton1993}}. }

So we take our theory to have models that are an $n$-tuple of a 4-dimensional manifold, a metric on it, and various fields. 
And for the sake of reporting Einstein, and setting up our later discussion, we shall take models very simply, as a triple $(M,g,T)$, where $M$ is a manifold, $g$ is a Lorentzian metric and $T$ is a rank-2 tensor representing stress-energy.  (We shall also often drop the $T$, and thus consider vacuum solutions. This does not affect any of the arguments to follow, and indeed is common in the hole argument literature.)  But as we said, the argument will not depend on the theory’s exact equations, or even on its field content.  

We will formulate general covariance, not as about the equations of the theory remaining the same under arbitrary coordinate transformations, but as about an arbitrary active transformation of the manifold---a diffeomorphism $d: M \rightarrow M$---producing, from a given model that satisfies the equations of motion of the theory, another such model: namely, by dragging along the fields of the given model. 
 We will think of dragging along a field as pushing it forward.  For since diffeomorphisms are smooth bijections with a smooth inverse, the differences between pulling back, and pushing forward, a geometric object such as a field, will not matter in this paper. So we will just write $d^*$ for the drag-along by the diffeomorphism $d$.

So we write general covariance, in terms of models of the theory, as follows:
\begin{quote} If ${\bf M}_1 := (M,g,T)$ is a model, and $d:M \rightarrow M$ is a diffeomorphism, then:\\
 ${\bf M}_2 := (d(M), d^*(g), d^*(T))$ is also a model.\footnote{This is sometimes called `active general covariance’. If one adopts this jargon, then the requirement that a theory be formulated so as to admit arbitrary coordinate transformations is correspondingly called `passive general covariance’.}
\end{quote} 
Clearly, two such models ${\bf M}_1$ and ${\bf M}_2$---with $d$ not the identity map---will in general differ as to how the metrical and material properties and relations encoded by $g$ and $T$ are distributed over  the points of the manifold. In this Section’s exposition of the hole argument, we will take these manifold points to be spacetime points.  So what we have just said implies that the models will in general differ over which spacetime points have which metrical and material properties and relations. (But later,  especially from Section \ref{sssec:metlesslm}, e.g. footnotes \ref{mathlpt} and  \ref{mathlpt2}, we will be more careful about the distinction between manifold points (which might be pure-mathematical entities) and the physical spacetime points they represent.) That is, the models will in general differ about such matters, when points  are taken to correspond, or be identical, according to the identity map on $M$. For example, if $d(p) = q$ (i.e. $p, q \in M$, and $d: p \mapsto q$), and $p$ has scalar curvature 5 in ${\bf M}_1$,  then: in ${\bf M}_2, q$ has scalar curvature 5---and $p$ has as its scalar curvature whatever is the scalar curvature, say 7, of the point $d^{-1}(p)$ in ${\bf M}_1$. 

There is a 
comment here, that is important since it has caused confusion about the interpretative, or philosophical, issues; (cf. footnote \ref{TushJamesVsHM} in Section \ref{subsec:31notdissolve}). Namely: we said that for $d \neq Id_M$, the models ${\bf M}_1$ and ${\bf M}_2$ will {\em in general} differ as to how the properties and relations are distributed over the spacetime points. The qualification `in general’ signals the fact that (for almost any theory), some models ${\bf M}_1$ have an {\em automorphism}, i.e. a non-identity diffeomorphism $d$ of their manifold such that the dragged-along model $(d(M), d^*(g), d^*(T))$ is exactly the same as ${\bf M}_1$. In other words: $d$ drags $g$ and $T$ into (coincidence with) themselves (respectively). So although $d: p \mapsto q$, with $q\neq p$, nevertheless $p$’s array of metrical and material properties and relations---in philosophical jargon: $p$’s {\em qualitative profile}---is the very same, in this selfsame model ${\bf M}_1$, as $q$’s qualitative profile.\footnote{\label{doublet}{We might write $p$’s qualitative profile, in the model ${\bf M}_1 := (M,g,T)$, in an obvious notation, as a doublet $(g|_p, T|_p)$. Then {\em in general} the image point in a dragged-along model has the same profile as the argument point in the given ``domain” model. That is: if $d(p) = q$, i.e. $d: p \mapsto q$, then $q$’s profile in the dragged-along model $(d(M), d^*(g), d^*(T))$ is the doublet $(d^*(g)|_{d(p)}, d^*(T)|_{d(p)})$. But if $d$ is an automorphism of $(M,g,T)$, then $(d(M), d^*(g), d^*(T)) = (d(M),g,T)$; and so if $d(p) = q$, then $q$’s profile is the same as $p$’s: the doublet $(d^*(g)|_{d(p)}, d^*(T)|_{d(p)}) = (g|_p, T|_p$).}}

 This point is also sometimes put in terms of distinguishing two senses of the word `isometry’ (or more generally, of `isomorphism’), and using the jargon of category theory. Thus for our case of Lorentzian manifolds (now for simplicity setting aside $T$), one says: given a  category whose objects are pairs $(M,g)$ of smooth manifolds $M$ and Lorentzian metrics on $M$, and whose arrows are isometries, where an \emph{isometry} $d$ from $(M,g)$ to $(N, h)$ is a bijection  between $M$  and $N$, such that $d$ and its inverse are smooth and also $d^*g=h$. This includes cases where we assume the underlying manifolds are \emph{identical} as sets, $M=N$. Of course, two isometric metrics, with $g\neq h$, can share the same base space. An \emph{automorphism} in this category consists of such an isometry with $d$ from $(M,g)$ to itself, i.e. such that $d\in$ Diff$(M)$ and $d^*g=g$.

Setting aside this comment about automorphisms, and returning now to the generic situation in which ${\bf M}_1$ and ${\bf M}_2$ differ as to how the properties and relations encoded by $g$ and $T$ are distributed over the spacetime points: let us finish our rendition of the hole argument. 

First, the hole argument now makes another assumption, additional to general covariance. It is about representation i.e. about how models represent physical possibilities. Namely, for what we just called the generic situation for ${\bf M}_1$ and ${\bf M}_2$, the argument assumes:
\begin{quote}
({\em Distinct}): If $d$ is the identity map except on a small patch $H \subset M$ (`the hole’), then ${\bf M}_2$ is, or represents, a different physical possibility than does ${\bf M}_1$.
\end{quote}
The punchline is now immediate. General covariance and ({\em Distinct}) imply that ${\bf M}_1$ and ${\bf M}_2$ are, or represent, different possibilities, although they exactly match each other outside $H$.   So the theory is indeterministic. But indeterminism is unacceptable. Hence, Einstein concluded, general covariance must be given up. Of course, with the later triumph of general relativity, 
({\em Distinct})   became the suspect, or culprit, in the argument; (as suggested by the idea of `Sophistication' at the end of Section 1.1,   and as we will see in Sections 2 and 3).\footnote{\label{Einstein}{As we admitted, we have formulated the hole argument in anachronistically modern terms. This will not affect later Sections. But we should note that for Einstein in 1913, ({\em Distinct}) was an implicit assumption: and when after November 1915, he returned to re-assess the argument,  he of course denied ({\em Distinct}). More specifically, he developed what   has come to be called   `the point-coincidence argument’, to convey the idea that which point has which qualitative profile of metric and material properties is what physicists now call `gauge'. Cf. \cite{Giovanelli2021}.   }}

Two final remarks, both limiting the scope of what follows. The first is about ({\em Distinct}), the second about  determinism. 

(1): In  ({\em Distinct}), the words `is, or represents' signal the philosophical debate about how models and physical possibilities are related to each other; (and the further metaphysical debate about what physical possibilities---philosophers' `possible worlds'---really are). Although we will glimpse these debates, in Sections \ref{sec:2debate} and \ref{sec:3dissolve?}, we will not need to take a stand on them.

 (2):  Determinism is a large subject: (cf. e.g. \cite{Earman_det, Earman_world}).
But we will only need the following claims, which are agreed by all hands.  The intuitive idea of determinism is of course that a theory is deterministic if among its models (or more metaphysically: the possible worlds they represent) two models matching exactly on the state of some appropriate region of spacetime (in many formulations: a time-slice, or all history up to a time slice) implies that they match throughout spacetime. So if `exact matching’ must include the facts about which point has which qualitative profile  (facts that philosophers call `non-qualitative', because  they concern {\em which} point), then a pair of models like ${\bf M}_1$ and ${\bf M}_2$ in the hole argument are undoubtedly a counterexample to this intuitive idea   (i.e. to the implication just formulated). That is:  they illustrate indeterminism. Besides, if we take indeterminism to require the explicitly metaphysical idea of two different possibilities, then assuming ({\em Distinct}), such models are, or represent, different possibilities. Furthermore, the larger the region on which the models match exactly, i.e. the smaller the ``hole” $H \subset M$ on which they disagree,  the stronger is the violation of determinism. (For a statement of implication is logically weaker, when its antecedent is logically stronger.) 
  
  Agreed: there is plenty more to say---both in physics and in philosophy---about the formulation and evaluation of determinism, including for general relativity. And again, we will glimpse some of these issues later (cf. especially (3), in Section \ref{sssec:cpart}). But this paper will not need to take a stand on them.

\section{Act 2: 1987 to 2000: Philosophical debate}\label{sec:2debate}
With the growing success of general relativity, the hole argument was mostly forgotten within physics. But within philosophy, interest in it revived in the late 1980s: very largely on account of \cite{EarmanNorton1987}.\footnote{Although  Earman and Norton's paper spawned a large literature, we say `very largely', not `exclusively', because some other contemporary discussions---including previous papers by Earman and by Norton, and Earman's book \cite{Earman_world}---also influenced philosophers. A significant recent analysis of that literature, and its current importance, is \cite{Weatherall2020}. Note also that Earman and Norton's philosophical writings were prompted in part by earlier historical work, by authors such as Stachel and Norton (cited in footnotes \ref{history} and \ref{history2}).}    In this philosophical revival, most of the attention has been on what Section \ref{ssecEin13} labelled ({\em Distinct}), rather than the argument’s other ingredients, general covariance and determinism. That is unsurprising in so far as ({\em Distinct}) is about interpreting the theory---philosophers’ {\em metier}---while general covariance and determinism seem established features of the theory.\footnote{Here, `seem established’ is intended to describe the attitude of most {\em philosophers}. For agreed: within physics, there are both rich traditions and ongoing efforts about: (i) how to better understand general covariance (also known as: diffeomorphism-invariance), and connected issues like the definition of observables, especially in connection with the hunt for a quantum theory of gravity; and (ii) definitions of, and theorems about, determinism in general relativity. Indeed, nowadays most physicists
 take the current interest of the hole argument to be just its directing one’s attention to topics (i) and (ii).  In the sequel, we will glimpse these topics.}

The distinctive, and perhaps surprising, feature of this philosophical revival was to articulate a new ``ingredient" of the argument: a philosophical doctrine called {\em `substantivalism'}. Thus Earman and Norton cast it as an argument against, not just ({\em Distinct}), but also substantivalism (Section \ref{ssec:revival}). And from 1987 to about 2000, other philosophers’ replies to them were mostly defences of substantivalism  (Section \ref{ssec:3replies}).

\subsection{1987: Earman and Norton against substantivalism}\label{ssec:revival}

 As we mentioned in Section  \ref{ssec2aims},  `substantivalism'  is vague: the broad idea being that spacetime is, or is represented by, a manifold, and exists independently of its material content. But we will take it (as we think Earman and Norton, and the ensuing literature, mostly do) as the claim that spacetime points are {\em objects}. (We will shortly be more careful about the distinction, signalled by our phrase `or is represented by’, between spacetime points and manifold points.) Three clarifications of this claim (in descending importance) are in order.

 (1): Beware: `object' is here meant in the ``thin'' or logically weak sense of modern logic, viz. as the referent of a singular term, which is therefore subject to   statements of identity and non-identity. So this sense, which derives especially from the work of Frege (and was later entrenched by the large influence of Quine), abjures the ``thick'' or logically strong (and so controversial) connotations of the traditional philosophical (in particular, Aristotelian) word `substance'. 

(2): Beware: confronted with this definition of `substantivalism', a physicist---indeed, anyone who is duly modest about what we can now know, or reasonably believe, about the ultimate micro-structure of spacetime---is likely to demur. They will say: `How could anyone claim to know either way?' But there is a distinction. The debate here is about how best to interpret general relativity (and maybe other spacetime theories): not---sorry, O physicist!---about reality itself. That is: the debate aims to settle what general relativity (and maybe other theories) {\em say} the world is like, thus of course leaving it to scientific enquiry---ultimately experiments---to settle whether the world is really as general relativity etc., so interpreted, say it is. So really `substantivalism'---in this literature, and in this paper---claims that {\em general relativity should be interpreted as committed to}   spacetime points being objects.\footnote{Agreed: there is  controversy even about the distinction we have invoked.   Thus some philosophers, notably van Fraassen  (e.g. \cite{Bas1980}), argue that even if a theory had complete experimental success, and satisfied all our other evaluative criteria (such as being conceptually unified, meshing with our other successful theories etc.), we could rationally remain agnostic concerning everything the theory says about matters lying beyond the experimental data. But throughout this paper, we set aside these general questions about exactly what beliefs about reality, a theory's complete success (experimental and otherwise) mandates. Trying to be practical and in the spirit of physics, we say: let us worry about such questions if and  when  we get to be so lucky!}   

(3): Like the literature, we have so far talked always about spacetime points, not regions. But we should clarify that this is for expository convenience, and not because substantivalists are in any way reluctant to say that regions are objects, just as much as points are. (Here, `say that' must of course be read as `say that general relativity says that'; cf. (2) above.) The reason is that modern logic makes light work of the passage from commitment to points, and commitment to regions. For it construes regions either as sets of points, or as mereological fusions of them.\footnote{`Mereology' is   modern logic's name for the study of the part-whole relation; and a {\em fusion} of objects is, intuitively, the aggregate of them. More formally, it is the whole (i) that has them as parts, and (ii) whose every part overlaps at least one of the objects. As this explanation of `fusion’ suggests, mereology is in practice a lot like set theory---but without the empty set.} 
 And logicians and philosophers tend to accept as unproblematic both the operation of forming the set of some given objects, and the operation of forming their fusion. So on both approaches, a theory's being committed to points is taken to imply its being committed to regions.

With substantivalism thus clarified, it is easy to guess how it enters in to Earman and Norton’s version of the hole argument. The main point is that they take substantivalism to  imply
({\em Distinct}).\footnote{\label{acid}{Their jargon is different for ours. Their `Leibniz equivalence’ states that isomorphic models represent the same physical possibility (cf. our `Sophistication’): which they call the `acid test of substantivalism’, i.e. which, they say, substantivalists must deny \cite[pp. 521-522]{EarmanNorton1987}.}} 

That is undoubtedly plausible: in effect, as a corollary of the meaning of `object'. Thus consider two distinct objects (which need not be spacetime points), call them $a$ and $b$; and two distinct arrays of properties and relations (what Section \ref{ssecEin13} called `qualitative profiles') such objects can have, call them $F$ and $G$. Then surely $a$ having $F$ and $b$ having $G$ is a different possibility than $a$ having $G$ and $b$ having $F$.\footnote{\label{Alice} And this is so even if the arrays $F$ and $G$ are exhaustive, i.e. encode all of the objects’ properties and relations. This conviction, that there are two possibilities here, is compelling and recurs in the literature. For example, it is the moral of the parable of Alice and Barbara in \cite[end of Section 3]{Pooley_Read}’s critique of \cite{Weatherall_hole}: to which we will return in Section \ref{subsec:31notdissolve}.}   Or recall Section 1.2's example of a diffeomorphism $d: M \rightarrow M$ where $d(p) = q$. We said that $p$ has scalar curvature 5 in ${\bf M}_1$, so that in ${\bf M}_2, q$ has scalar curvature 5, and $p$ has as its scalar curvature whatever is the scalar curvature, say 7, in ${\bf M}_1$ of  $d^{-1}(p)$).  If points are objects, surely these statements should be taken at face-value as regards what are the represented possibilities. In particular: the threat that $p$ has contradictory properties (having curvature 5, and having  curvature 7) is most straightforwardly answered by saying that ${\bf M}_1$ and ${\bf M}_2$ are or represent different possibilities, i.e. by endorsing ({\em Distinct}).

Once one assumes that substantivalism implies ({\em Distinct}), the argument is clear. There is a dilemma: one can maintain that general relativity is deterministic, by denying ({\em Distinct}) and so also denying substantivalism; or one can be a substantivalist, be committed to ({\em Distinct}), and say that general relativity is indeterministic. 

Then Earman and Norton go on to urge the first option. Though (like Einstein!) they give no precise definition of determinism, they emphasise that their favouring the first option is {\em not} because of any conviction that general relativity (or any other physical theory) must at all costs be classified as deterministic. They are entirely open to the idea that a physical theory is indeterministic. But, they say, if determinism fails `it should fail for a reason of physics’ (ibid. p. 524). And this is their objection to the dilemma’s second option: it makes general relativity indeterministic for a distinctively metaphysical reason that, they say, should have no weight  in classifying physical theories as indeterministic.

\subsection{1987 to 2000: Defending substantivalism}\label{ssec:3replies}
 For the most part, philosophers have replied to Earman and Norton's dilemma by defending substantivalism. And they took this to involve: denying that substantivalism implies ({\em Distinct}). We agreed at the end of Section \ref{ssec:revival} that this implication is undoubtedly plausible. For if spacetime points are objects, then surely swapping which point has which qualitative profile of field-values makes a genuine difference? Yet maybe this implication can be denied---albeit at the cost of what boxers call `fancy footwork'.  So this Section is about such philosophical fancy footwork. (At the end of this Subsection, we will return to the tactic of defending substantivalism by accepting its implying ({\em Distinct}), and just admitting that general relativity is indeterministic.) 

Besides, some of these replies went further than denying ({\em Distinct}): which in our formulation is only about general relativity, viz. any model ${\bf M}_1$ and its drag-along ${\bf M}_2$. For recall from (1) at the end of Section \ref{ssec2aims} the doctrine, nowadays called {\em Sophistication}. Namely: for any theory---not just general relativity---whose models are or represent possibilities:  {\em any two isomorphic models are, or represent, the same physical possibility}.\footnote{  Section \ref{subsec:31notdissolve} will address subtleties about the context of such representations, that are articulated in more recent literature.}  Thus some of these replies endorsed (or came close to endorsing) {\em Sophistication};  (though the label `sophistication' was only common after 2000).
So in briefly reporting some of these replies (and in the next Section), Sophistication will be a theme. 

For our purposes, we can group the denials of ({\em Distinct}) under three headings, which we call:  {\em metrical essentialism}, {\em the drag-along response} and {\em counterparts}. The second and third of these will be (different) responses to the difficulties faced by the first; the third---which we favour---also responds to a difficulty faced by the second.  The second will be tantamount to endorsing Sophistication for general relativity (but not necessarily for other theories).\footnote{As to our labels for these responses, the second is our invention; but the first and third are established jargon in philosophy.   For further references, cf. below: especially Sections \ref{sssec:dar} and \ref{subsec:31notdissolve} for the second response, and Section \ref{sssec:cpart} for the third.}

We should clarify at the outset that although we will here criticise the drag-along response, we are in fact sympathetic to Sophistication. So we do not intend `fancy footwork' as a slur. This is not least because (as mentioned in (2) at the end of Section \ref{ssec2aims}) one of us (HG) has urged parallel reasons for it, in general relativity and in gauge theories. But in this paper, we will not argue directly for Sophistication, even just for general relativity. Nor will the framework (for both general relativity and gauge theories) that we propose in our second paper, Part II, add to the {\em philosophical} case for Sophistication. For that framework's merits lie in treating non-isomorphic spacetimes (in gauge theory: non-isomorphic field-configurations); and thus in placing discussion of Sophistication in a wider context.

\subsubsection{Metrical essentialism}\label{sssec:metlesslm}
`Essentialism’ is jargon for the philosophical idea that some objects (maybe all objects) have one or more attributes (i.e. properties or relations) that they could not possibly lack. That is: in any possible situation in which the object exists, it has the attribute in question.

The traditional examples are of properties that are natural kinds, in particular an organism’s species. (The tradition goes back to Aristotle.) Thus a human, say Socrates, could not possibly be non-human, e.g. a frog; this oak tree could not possibly be an elm tree; and so on. But one can envisage relational examples, even everyday or biological ones. For example, the parent-child relation: this child could not possibly lack their actual parentage---different parents would imply that the child is a different person. (So in the imagined possibility, `the child’ does not refer to the actual child.) 

This idea immediately suggests a way to deny ({\em Distinct}). For suppose that a given model ${\bf M}_1$ is, or represents, a possibility: a `way that a general-relativistic world could be’. And suppose that some metric or material attribute (property or relation) of a point $p$ in the hole $H \subset M$ that is shifted by the diffeomorphism $d$, i.e. $d(p) \neq p$, is both:\\
\indent \indent (i) essential to $p$, and (ii) not possessed by $d^{-1}(p)$.\footnote{\label{mathlpt}If one takes the points of the manifold $M$ to be in some sense pure-mathematical points, not genuine spacetime points, then it will suffice to suppose that some such attribute, represented by a field-value of $p$ (cf. footnote \ref{doublet}) is (i) essential to the spacetime point that $p$ represents, and (ii) not possessed by the spacetime point that $d^{-1}(p)$ represents. And this supposition seems compulsory. For after all, we supposed that ${\bf M}_1$ represents a possibility, even if it is not identical to one; and it surely does so by its manifold points representing spacetime points.}\\
It follows that ${\bf M}_2$---where $p$, as an element of the range of $d$, ``inherits” the qualitative profile possessed in ${\bf M}_1$ by $d^{-1}(p)$, and has ``bequeathed” to $d(p)$ the qualitative profile it possessed in ${\bf M}_1$---does {\em not} represent a possibility. For in ${\bf M}_2$, $p$’s qualitative profile lacks an attribute essential to $p$ (the attribute having been ``bequeathed”, in this model, to $d(p)$). In short: ${\bf M}_2$ fails to be, or represent, a possibility, because it is faithless to $p$'s essential attribute. ${\bf M}_2$ ``paints $p$ with the colours” of $d^{-1}(p)$ (i.e. of $d^{-1}(p)$, within ${\bf M}_1$), and at least one of those colours is for $p$ an impossible colour.\footnote{\label{mathlpt2} To make the wording simpler, we have stated the argument taking $p$ etc. as spacetime points with attributes. But if one takes the points of the manifold $M$ to be pure-mathematical and only represent spacetime points, the argument can be rephrased, though at the price of cumbersome language, along the lines of  footnote \ref {mathlpt}.}  

Note that we have given this argument in terms of $p$ both (i) ``losing an essential property to $d(p)$”, and (ii) ``not re-acquiring it from $d^{-1}(p)$”, as our attention passes from ${\bf M}_1$ to ${\bf M}_2$. But of course, we could equally well have chosen $d(p)$ as ``the victim of loss”. We could have supposed that $d(p)$ has, within ${\bf M}_1$, an attribute that is: (i) essential to $d(p)$, but (ii) not possessed by $p$; and so is, within ${\bf M}_2$,  (i') bequeathed to $d^2(p)$ but (ii’) not re-acquired from $p$.\footnote{  To be vivid, we have used metaphors of  loss, inheritance, colours etc. But it is clear that the argument could be made more formal, using pullbacks, and the doublets mentioned in footnote \ref{doublet}.}

 So much for the ``mechanics” of this denial of ({\em Distinct}). The question now is: for what attributes is it plausible to claim that they are both (i) essential to spacetime points, and yet also (ii) re-assigned among points by dragging along with a diffeomorphism? 
 
 The reply has been: metrical attributes---hence the label {\em metrical essentialism}. These attributes are typically not properties, but rather relations. So they are encoded, not by scalars like the scalar curvature, but by vectors and tensors, like the metric tensor and numerical quantities derived from such tensors, like the length of a line between two points. But our argument allowed for essential relations as well as properties. So the way is open to deny ({\em Distinct}) by holding, for example, that  the metrical relations of a spacetime point $p$ to other, maybe to all other, points (for example: the lengths of lines from $p$ to these other points) are essential to $p$.\footnote{\label{FnPointil}{For the interplay between (a) philosophical notions of property (including intrinsic and extrinsic properties) and relation, and (b) geometry’s and mechanics’ notions of scalar, vector, tensor etc, cf. \cite[Section 2.1.1]{ButterfieldPoint}.}}    

The first advocate of this reply to Earman and Norton was \cite{Maudlin_substance}. He invokes Aristotle in defence of the general idea of essential attributes, and he invokes Newton’s Scholium to the Definitions in {\em Principia} in defence of spacetime points (for Newton: temporal instants and spatial points) having their metrical relations to each other essentially (p. 544).

But despite this noble pedigree, this reply has a severe limitation, due to general relativity’s having non-isometric models. For most advocates of essential attributes (from Aristotle onwards) hold the view for {\em actual} objects, and are ``modestly” silent about whether merely possible objects---of course, a unicorn is a standard example---have essential attributes. Since the debate here concerns how best to interpret general relativity, assuming it is true (cf. comment (2) in Section \ref{ssec:revival}), the modest version of this reply would confine its claims to the (putatively) actual spacetime points, i.e. the points of the actual general relativistic cosmos. But then it says nothing about the identity of points in spacetimes non-isometric to the actual cosmos, except that none of such a spacetime’s points is the same point as any actual point. (For thanks to the non-isometry, any of these points lacks at least one of the intricate web of metrical relations that the reply maintains are essential to the actual points.) In particular, it says nothing about how to reply to the hole argument applied to such a spacetime. So the challenge of the hole argument remains.

We do not claim this is knock-out objection against metrical essentialism: as always in philosophy, it can give some reply.  But there are good reasons to think its prospects are bleak; (as argued by e.g. \cite[Section 5]{Butterfield_hole} and \cite[Section 4]{Pooley_routledge}).\footnote{We will not go in to details. We just note that it does not help essentialism either (i) to   relativize essential attributes to isometry classes or (ii) to advocate a different set of attributes than the metrical ones as essential to points. For even admitting (i) and (ii), essentialism  is still silent about which $\bf M$ along an isometry class should be the one whose points bear the essential attributes, and a priori all such models seem equally metaphysically possible. Another worry is that metric essentialism has difficulty expressing counterfactuals such as "if the mass of the sun was doubled, spacetime would be more curved around it".  One merit of our preferred response via counterparts in Part II is that it answers these worries. We also note that although we have formulated this objection by speaking of points in spacetimes, and so in effect by taking models ${\bf M}_1$ etc. to {\em be} possibilities, not just to represent them: nevertheless the objection could be stated while respecting such subtleties (albeit more cumbersomely) along the lines of footnotes \ref{mathlpt} and \ref{mathlpt2}.}

\subsubsection{The drag-along response}\label{sssec:dar}
 To avoid the difficulties of metrical essentialism, it is natural to try and deny ({\em Distinct}) by saying that ${\bf M}_2$ is, or represents, the very same possibility as ${\bf M}_1$---despite its difference from ${\bf M}_1$ about which point has which qualitative profile. The basic idea, in philosophical jargon, is that points are {\em individuated} by their pattern of qualitative properties and relations, as encoded in the metric and matter fields. (The jargon is that saying a set of objects $X$ `is individuated by'  a set of attributes $Y$ means that there are 
sufficient conditions, expressed in terms of $Y$, for statements of identity about members of $X$.)  But the drag-along response takes this idea further, asserting that such individuation also applies ``between'' isomorphic models.\footnote{\label{Komar}{We say `takes the idea further' to signal the fact that one can take the attributes $Y$ to individuate the $X$s, within a  model or possibility, without  being committed to their doing so ``between" models, even isomorphic ones. In general relativity, this is illustrated by  Komar observables (cf. \cite{Komar_inv}). The idea is that, at least in a generic {\em non}-symmetric  spacetime, no two points have the same set of values of certain observables e.g. scalar curvature; so that the points can be labelled, i.e. uniquely referred to, by their values. But this intra-model individuation by the observables does not imply that between two models, two points must be identified by their having the same values---which is what the drag-along response will claim.}}

Thus suppose again (cf. the start of Section \ref{sssec:metlesslm}) that the given model ${\bf M}_1$ is, or represents, a possibility. And suppose that some point $p$ in the hole $H \subset M$ is shifted by $d$ to a distinct point,  $q \equiv d(p) \neq p$, that {\em within ${\bf M}_1$} has a different qualitative profile (some different metric or material attribute) than $p$.  Then the proposed reply to the hole argument
 is that, {\em within ${\bf M}_2$},  we should ``rebrand" $d(p)$ as ``really being" $p$; or in other words, ``replace $d(p)$ with $p$". 
 
It is this proposed reply that we call {\em the drag-along response}. The label is meant to suggest that a diffeomorphism ``drags along the identity’’ of each point in its domain-model to its codomain-model, so that the identity ``follows" the qualitative profile.\footnote{ So the contrast with the previous Subsection is this: whereas metrical essentialism condemns ${\bf M}_2$ for representing no possibility, because it is faithless to points’ essential attributes, the drag-along response takes it to represent the very same possibility as ${\bf M}_1$---by using attributes, i.e. qualitative profiles, to ``rebrand” points.} 
 
 For this paper’s purposes, we need to make just three comments, (1)-(3), about the drag-along response. (1) is about making it precise; (2)  is about our own attitude to it; and (3)  present problems for it.
 
 \medskip
 
(1):  {\em Making the response precise}:--- Of course, talk of `rebranding’ or `replacing’ points, or ``dragging identity”, is metaphorical. And to make the proposal precise, it is of course  not enough to simply replace, in the mathematics, the non-identity function $d$ by the identity function $Id$ on $M$. Doing only that would simply change the subject, and ignore the non-mathematical  interpretative issues raised by the hole argument, i.e. by the given function $d$.

But the proposal has been made precise, namely by \cite{Stachel_Iftime_short, Stachel_Iftime_long};   cf. also \cite{Stachel1986} and \cite[Section 4]{Stachel_lrr}. 
They postulate a fibre bundle for which they can {\em define} a spacetime point as, in effect, a set of field-values (what we have called a `qualitative profile'). This means that there is simply no issue about relating   points between isomorphic models: the points follow their qualitative profiles---implementing the drag-along response.  

But we need not go into details about these definitions, even though this approach, like our own in Part II, uses fibre bundle ideas. For the approaches are very different. As we announced in Section \ref{ssec2aims}, ours but not Stachel's addresses how to relate   points belonging to {\em non}-isomorphic models; and it is less abstract, with specific examples that can be worked out.

More generally, we need not try to make the drag-along response precise. For our purposes, it suffices to say that it has two components:  (a) any two isomorphic models of general relativity represent the same possibility (i.e. Sophistication for general relativity); and  (b) this is compatible with substantivalism, since points are individuated by their qualitative profiles.\footnote{\label{Hartry}{ Among the many formulations of the drag-along response to be found in the literature, here is a clear one expressing both these components: ` \textit{“individuation of objects across possible worlds” is sufficiently tied to their
qualitative characteristics so that if there is a unique 1-1 correspondence
between the space-time of world A and the space-time of world B that
preserves all geometric properties and relations (including geometric relations
among the regions, and occupancy properties like being occupied by
a round red object), then it makes no sense to suppose that identification of
space-time regions across these worlds goes via anything other than this isomorphism.}' \cite[p. 77]{Field_soph}.}}  

 \medskip
 
(2):  {\em Our attitude}:--- Our own view about the drag-along response is fourfold.\\
\indent \indent (a): It is plausible, since in many examples its verdicts about ``which point is which” are right (as the quotation in footnote \ref{Hartry} says). \\
\indent \indent (b): But it faces   problems. In (3) below, we report one: about models that have an automorphism.  \\
\indent \indent (c): As we will argue in Section \ref{sec:3dissolve?}: even for models that have no automorphisms, its verdicts are not always right. So it is certainly not compulsory  as a matter of mathematics, as some authors have urged.\\
\indent \indent (d): However, there is a grain of truth in the drag-along response, and even in the claim that it is mathematically compulsory. For at the end of our second paper, i.e. Part II, we will see that for a model that has no automorphisms, and an isomorphic copy of it related by a diffeomorphism $d$, our fibre bundle framework picks out the map $d$ as the preferred way to relate the models.\footnote{(i): For automorphisms, cf. the comment after Section \ref{ssecEin13}’s formulation of general covariance. (ii): Note that in Part II’s discussion, the map $d$ is written as $g$, since it an element of the {\em group} Diff($M$).
}
 
 \medskip
 
 (3):  {\em Problems}:--- The drag-along response faces problems. We pick out one that is easily explained and has been discussed in the philosophical literature.\footnote{\label{Belot} {Another problem is more intricate and technical. It was introduced to the philosophical literature (so far as we know) by \cite{Belot50}, especially Sections 4.3 and 4.4, pages 964-970. In short, the problem is that, for spacetimes that are asymptotically flat at infinity, it is standard practice to treat some isomorphisms (for vacuum spacetimes: some isometries) as generating from a given model (in the same manner as the hole argument) a new physical possibility, that differs from the given model by a time-translation at spatial infinity. Obviously, this standard practice threatens Sophistication about general relativity, and thus the drag-along response. The topic is intricate and far from settled: for instance, in \cite{Samediff_1b}, the standard practice is reconstrued so as not to threaten Sophistication.  But we cannot here pursue this topic; though we will briefly return to it in Section \ref{subsec:31notdissolve}.}} It is about {\em symmetric} models, i.e. models that have an automorphism (additional of course to the identity map on the manifold). In short, the drag-along response is led to make contradictory assertions about the identity of some points in any such model. 
For suppose a model ${\bf M} := (M,g,T)$ has an automorphism $d$ with $d: p \mapsto q \in M$ and $q \neq p$. Then the drag-along response's idea of rebranding or replacing points leads it to say that $q$ ``is officially” $p$. Besides, the problem can also be stated without depending on a single manifold $M$. Given such a model $\bf M$ with symmetry $d$ (with $d(p) = q \neq p$), there will be other models ${\bf M’}$ with a distinct manifold $M’$, with two isomorphisms from ${\bf M’}$ to ${\bf M}$, one of them obtained by composing the other with $d$. That is: there are isomorphisms $d_1: M’ \rightarrow M$, $d_2: M’ \rightarrow M$, and a point $r \in M’$ with  $d_1(r) = p$ and $d_2 := d \circ d_1$ so that $d_2(r) = q$.  In such a case, $p$ and $q$ are equally good candidates to ``officially be” $r$---and so be identical to each other. But since $q \neq p$, the drag-along response seems committed to contradiction. Furthermore, the trouble can be, so to speak, everywhere in the model. That is: if the model $\bf M$ is homogeneous, i.e. for any two points $p, q \in M$, there is an automorphism with $p \mapsto q$, then any two points in $M$ get officially identified by the drag-along response---so that officially there is just one spacetime point.\footnote{This problem has a previous incarnation, in the different though related debate about the philosophical doctrine called `spacetime structuralism’: which is, roughly, the view that spacetime points can be individuated by what we have called their qualitative profiles. \cite{Wuthrich_abysmal}  presented the problem for homogeneous models, as a refutation of spacetime structuralism, calling it an `abysmal embarrassment’. \cite{Muller_abysmal}  is a reply, based on views about individuation and identity developed by him and Saunders.}

\subsubsection{Counterparts}\label{sssec:cpart}
 We have just seen that metrical essentialism has difficulties about non-actual points, and that the drag-along response has difficulties about two or more points being equally good candidates for being identical to a given point. Both sorts of difficulty are avoided by a framework, called {\em counterpart theory}, for treating the identity of objects (all sorts of objects, not just points) in different possibilities; and we now review this.

We must again be brief. For counterpart theory is a large subject in the logic and metaphysics of possibility; and even within the smaller philosophy of physics literature, its application to spacetime points is a central topic in discussion of the hole argument. With an eye on what we will need in Section \ref{sec:3dissolve?} and Part II, we shall confine ourselves to three comments, (1) to (3). (1) introduces the main ideas as they are set out in the purely philosophical literature. (2) describes the application to spacetime points and in answer to the hole argument.  Finally, (3) returns us to the topic of determinism.

\medskip

(1): {\em The main ideas}:--- Counterpart theory was invented and developed by the great metaphysician David Lewis (mainly in his \citep{Lewis1968}, \cite[p. 38-43]{Lewis1973}  and  \cite[Ch. 4]{Lewis1986}). Lewis is best known for advocating a framework of {\em possible worlds} (i.e. maximally specific possibilities) as the best way to address all sorts of philosophical problems  ranging from the nature of propositions to causation; and counterpart theory is part of that framework.\footnote{\label{DKL}{
Among philosophers, Lewis is also famous for a belief he called {\em modal realism}: that each non-actual world---i.e. each way the cosmos could possibly be though in fact it isn’t that way---is no less real than the actual world, i.e. cosmos, around us. That belief, almost all other philosophers (including us) reject as incredible---in the literal sense. We might have said `notorious’ rather than `famous’: though all agree that Lewis makes a very good case for his modal realism; initially in   \cite[Ch. 4.1]{Lewis1973}, and at length in \citep{Lewis1986}.

 But fortunately, all agree that counterpart theory can be detached from modal realism, so that from now on we will set aside the latter. And we also set aside the more popular rival accounts of the ontological status of possible worlds: of which  \citep{Lewis1986} gives a fair, indeed masterly, survey.}} 

The key claim of counterpart theory is that objects in different possible worlds are never {\em identical} to each other. Rather, they are {\em counterparts} of each other---better or worse counterparts, according to how similar they are. Here is one of Lewis’ early formulations:
\begin{quote} 
In general: something has for {\em counterparts} at a given world those things existing there that resemble it closely enough in important respects of intrinsic quality and extrinsic relations, and that resemble it no less closely than do other things existing there. Ordinarily something will have one counterpart or none at a given world, but ties in similarity may give it multiple counterparts. \cite[p. 39]{Lewis1973}.
\end{quote}
As Lewis shows (and as this quote suggests), the great advantage of counterpart theory over saying that an object in a possible world can be literally identical with an object in another, is {\em flexibility}—thanks to similarity being vague and coming in degrees. In particular, since two objects can be equally similar to a given one, counterpart theory easily makes sense of statements like `I could have been twins’. (Cf. the drag-along response’s problem with symmetric models, that we saw in comment (3) of Section \ref{sssec:dar}.)\footnote{\label{Hubert} Besides, Lewis rebuts the objection that when we say `Hubert Humphrey might have won the USA’s 1968 Presidential election’ (an  example that became famous in the philosophical literature), we mean that the actual Humphrey, ``our Humphrey”, might have won---not some distinct but appropriately similar person. Lewis agrees: indeed, that is what we mean. But the question is how another world can represent concerning the actual Humphrey, that he wins. And Lewis argues that counterparts do so: to use another philosophical jargon, they are suitable {\em truth-makers} for propositions about possibilities for actual objects, such as Humphrey \cite[p. 194-197]{Lewis1986}.

 Even setting aside counterpart theory, similarity is an important notion for various topics in foundations of physics. For example, similarity of models (solutions) is important for inter-theoretic reduction and approximation. \cite{FletcherWarsaw} is a fine survey, emphasising topological ideas---to which we will return in Section \ref{sssec:limits}.
}

\medskip

(2): {\em Counterparts of points}:--- So suppose we deny ``trans-world identity” for spacetime points in different possible worlds: they can only be counterparts. The first thing to say about this view is that it immediately implies  that ({\em Distinct}) is false. Thus this view affords a third way---additional to metrical essentialism and the drag-along response---to reply to Earman and Norton’s dilemma by upholding substantivalism, while denying that substantivalism implies ({\em Distinct}). (Recall the preamble to Section \ref{ssec:3replies}.)
   
The reason this view denies ({\em Distinct}) is analogous to, but simpler than, the reason that metrical essentialism denied it.\footnote{That reason is at the start of  Section \ref{sssec:metlesslm}.  Note the contrast with the reason that the drag-along response denies ({\em Distinct}): {\em it} takes the two models to represent the same possibility.} For if a given model ${\bf M}_1$ is, or represents, a possibility (a way that a general-relativistic world could be), then the model we called ${\bf M}_2$, i.e. the model obtained by dragging along the fields in ${\bf M}_1$, is {\em not} (nor represents) a different possibility. For it is based on the same manifold $M$ as ${\bf M}_1$; while on this view, none of the points in any such different possibility is a point in $M$.\footnote{\label{Frank}{This way to defend substantivalism in reply to Earman and Norton was first advocated by one of us (JB); \cite[p. 25-26]{Butterfield1987}, and at more length in  \cite[Sec. 6]{Butterfield_hole}. Since then, it has been elaborated. Three excellent recent papers are: \cite{Jacobs_reply, Cudek_counter1, Cudek_counter2}. Note also that to make the wording simpler, we here write as if spacetime points, whether actual or possible, are elements of a model’s manifold; as we did in Section \ref{sssec:metlesslm}. But as noted there (cf. footnotes \ref{mathlpt} and \ref{mathlpt2}) the argument can be rephrased without this assumption.}}

For the purposes of this paper and of Part II, there are two further aspects of our use of counterpart theory for spacetime points that we need to report. (Neither is contentious.) The first is  to clarify our use of the jargon `counterpart’, in relation to its use in general philosophy. The second is about determinism: which we take up in (3) below. 

We have said that counterparts are more or less similar, but never identical. And as the quotation from Lewis suggests: in his treatment of statements of possibility for ordinary objects, e.g. a person such as Humphrey, the relevant respects of similarity will be vague (though determined in part by the context in which the statement is made, e.g. the intentions of the speaker) and will typically involve   some balancing or trade-offs between various respects. But herein lies a contrast with our application of counterpart theory to spacetime points, as we have so far stated it, i.e. as part of a reply to the hole argument. 

For the respects of similarity that we invoke, in the precise setting of a spacetime theory such as general relativity, are not vague. They are---in our application as stated so far, and in the treatment of determinism in (3) below---a matter of exact matching of the points’ qualitative profiles of field-values. Here, exact matching is made precise by a diffeomorphism dragging along the geometric objects at a given point in the domain-model  
  into coincidence (respectively) with the geometric objects at the point’s image in the codomain-model. In particular, there is no need to strike a balance between various attributes, some shared and some not. It is solely a matter of the geometric objects being required to be preserved by the diffeomorphism.\footnote{Apart from scalar fields, the geometric objects or field-values will mostly encode extrinsic attributes. For discussion, cf. the reference in footnote \ref {FnPointil}.} 

This contrast is not a problem for us. After all, `counterpart’ is only a word, and an application of `counterpart theory’ to a specific topic, such as spacetime points or the hole argument, can perfectly well have a special feature, such as not being vague. But the contrast does prompt a clarifying philosophical remark: as follows. 

For there is another philosophical jargon (also introduced by Lewis) for objects that exactly match in this sort of way. Lewis calls such objects {\em duplicates}. But we prefer to retain the word `counterpart’ for three reasons; (in ascending order of importance, for us).\\
\indent \indent  (i): Lewis’ own use of `duplicate’ is tied to contentious doctrines of his (about what he calls `natural properties’): doctrines which, for reasons not relevant here, we do not endorse.\\
\indent \indent (ii): Being duplicates does not exclude being identical (neither in Lewis’ treatment of duplicates, nor in others’); while `counterpart’ has (thanks to Lewis’ large influence in philosophy) a firmly established connotation of excluding identity. And for our answer to the hole argument, the non-identity is important.\\
\indent \indent  (iii): For our work in Part II, viz. treating {\em non}-isomorphic spacetimes, regardless of answering the hole argument: exact matching of points will {\em not} be centre-stage. Instead, thanks to the non-isomorphism: Lewis’ ideas of degrees of similarity, and striking a balance between different respects of similarity, will apply  naturally to spacetime points, and will be centre-stage. And since these ideas were as much part of his original conception of counterparts as was the denial of ``trans-world” identity, it is  clearer for us to use the word `counterpart’, rather than e.g. `duplicate’.\footnote{For more discussion of (i) and (ii), cf. \cite[pp. 23-25]{Butterfield_hole}.  As to (iii), we will also see a natural use of the term `duplicate’ at the end of Part II. }

\medskip

(3): {\em Determinism revisited}:--- So far in this Section’s review, determinism has been a minor theme. At the end of Section \ref{ssec:revival}, we reported that Earman and Norton agree that determinism is not sacrosanct, but say that it should be given up only for reasons of physics, not for a philosophical reason such as defending substantivalism. Then this Subsection reviewed the dominant ``fancy-footwork'' reply to Earman and Norton, that substantivalism does not imply ({\em Distinct}); so that, the hole argument notwithstanding, it does not threaten determinism. 

But to adequately review this controversy, we should make two further comments about determinism: the first about philosophical strategy, the second about the definition of determinism.\footnote{ We hope to be adequate, but we must be brief. We set aside details not only about our two comments, but about other aspects of determinism in the philosophical literature: about which we recommend the three recent papers cited in footnote \ref{Frank}, Though brief, what follows is anyway enough for this paper and for Part II: in which details about determinism will not be center-stage.}

First: one need not agree with Earman and Norton that determinism should be given up only for reasons of physics. For `determinism’ is only a word: and physicists and philosophers could perfectly well agree to use it differently. (Recall that Earman and Norton give no precise definition of it.) That is: one can be a substantivalist in Earman and Norton’s sense, and agree with them that substantivalism implies ({\em Distinct}), and then respond to their argument by ``biting the bullet”. One simply concedes that general relativity is indeterministic, for the philosophical reasons about the identity of points that the hole argument reveals. But one then adds: `No worries. For general relativity as a physical theory need have---indeed, should have---no concern about such philosophical indeterminism.’  

This is undoubtedly a tenable response. But it highlights the need to formulate some precise definition (or definitions) of determinism, especially a definition that general relativity {\em does} satisfy. In terms of philosophical strategy, this need is especially pressing on Earman and Norton, who stress that they respect only `reasons of physics’ in evaluations of whether determinism holds.   But it hardly matters where the need is most pressing. For all parties must admit the benefit of precise definition(s) of a central term such as `determinism’.  This leads to our second comment.

Second: In the endeavour to give such a definition, one of course turns to the physics literature’s treatment of the existence and uniqueness of solutions to the initial-value problem for general relativity. In the philosophical literature about the hole argument, the first such effort was by one of us (\cite[p. 26-31]{Butterfield1987}; and \cite[Sec. 3]{Butterfield_hole}): focussing on the treatment by \cite[Ch. 7]{HawkingEllis}. In short, he argued for two claims---that meshed with the counterpart-theoretic reply to Earman and Norton. \\
\indent \indent \indent (i): There is a definition of determinism, labelled {\bf Dm1}, that {\em is} violated by general relativity. In effect, the reason it is violated is that it expresses determinism’s basic idea---that exact matching of two models on  the past implies exact matching  throughout the models---by having a single diffeomorphism, from the whole domain-model to the whole codomain-model, drag along the geometric objects into coincidence.\\
\indent \indent \indent (ii): But another definition of determinism, labelled {\bf Dm2}---that is obeyed by general relativity---can be extracted from the technical theorems expounded by Hawking and Ellis (especially that of \cite{ChoquetBruhat1969}).  Its main difference from {\bf Dm1} is that it expresses the regional-to-global implication (of exact matching), not with a single global diffeomorphism, but by allowing the implied global exact matching to be specified by a {\em different} diffeomorphism than the one that specified the assumed regional exact matching.
 This freedom to use a different diffeomorphism not only suggests the flexibility of counterpart theory. It also reflects the technical theorems’ allowance that one {\em development} (intuitively: one ``continuation into the future'') of given initial data can be an extension of another development---and this notion of extension includes cases where a diffeomorphism defines a new model by drag-along, as in the hole argument. 

But we should register here that the letter of (ii) was wrong, although the spirit was right: ({\em mea culpa!}). Butterfield’s ``transcription” from the technical theorems erred: though not in ways that affect our summary in (i) and (ii), or that matter for the hole argument and how to reply to it. The errors and how to amend them are stated very well in excellent papers by \cite[p. 8-9, and footnote 22]{Landsman2022} and \cite[Section 5.1]{Cudek_counter1}.\footnote{The main technical points, covered by both authors, are that {\bf Dm2} needs to be reformulated so as to be restricted to globally hyperbolic solutions to the vacuum Einstein equations, and to their maximal developments. Landsman also adds, more philosophically, that whereas both {\bf Dm1} and {\bf Dm2} {\em assume} the existence of a class of models, i.e. 4-dimensional solutions of the theory (as do all definitions of determinism in the philosophical literature), theorems solving an initial-value problem, like that of Choquet-Bruhat and Geroch (on which Landsman focusses), have the merit of {\em proving} that these solutions exist, given appropriate initial data.} In any case, these amendments are not needed in the sequel.

 \section{Act 3: 2016 to 2022: The hole argument dissolved? Threading spacetime points}\label{sec:3dissolve?}
 
So much by way of reviewing philosophers’ replies  during the 1990s to Earman and Norton’s revival of the hole argument. From about 2000, there was less discussion\footnote{But we recommend \cite{Pooley1998PhD} and \cite{Pooley_rel}.}, until about 2018: when \cite{Weatherall_hole} argued that mathematical practice rules out the hole argument. In short, he proposed that isomorphism is the mandatory standard for judging whether two points in (the manifolds of) two models of the theory correspond, or “are identical”. To put it in our jargon: the `drag-along response’ (cf. Section \ref{sssec:dar}) is compulsory, as a matter of {\em mathematics}, so that the hole argument is dissolved, or ``never gets started”.

This proposal revived debate. Some authors agreed, or broadly agreed, with Weatherall (e.g.  \cite{Fletcher_hole},  \cite{HalvorsonManchak_hole},  \cite{BradleyWeatherall_hole}   who also discuss precursors to the proposal). Others have disagreed (e.g. \cite{Arledge2019, Roberts_hole, Pooley_routledge, Pooley_Read, Roberts_disregarding, ReadMenon_hole}).

We ourselves disagree; and the goal of this Section is to give our reasons. Again, we must be brief: we cannot do justice to all the details of the recent literature.  We confine ourselves to stating what we consider the main reasons for disagreeing. They centre around the fact that mathematics and physics treat the identity of points (indeed, of objects in general) more flexibly than Weatherall et al. allow (Section \ref{subsec:31notdissolve}). This will lead in to our showing some examples of points in two models corresponding, or being ``identified", other than by isomorphism (Section \ref{subsec:32thread}). We propose to  call this the {\em threading} of points. Unsurprisingly, it will play a role in Part II's discussion of {\em non}-isomorphic spacetimes.       
 
 \subsection{Against dissolution}\label{subsec:31notdissolve}

  We have  already agreed that ``identifying" points in two models by an isomorphism, as in the drag-along response, is plausible; (cf. (2)(a) and footnote \ref{Hartry} in Section \ref{sssec:dar}) . But we deny that mathematical practice {\em dictates} this view; and that thereby, the hole argument ``never gets started’’.  
  
   We have three main reasons for this denial.  First, in (1) below, we disagree with specific claims of these authors. We hold that mathematical practice does not close off interpretative options. In particular, it does not refute ({\em Distinct}) in the ways they claim. Then in (2) we argue generally---independently of general relativity---that mathematical practice treats identity more flexibly than these authors allow. (1) and (2) will have a common moral: we should beware of trying to read philosophical morals directly off mathematics. Our third reason is again specific, but  more positive:  in some contexts, mathematics ``identifies" (or: puts in correspondence) points in two models other than by isomorphism---the next Subsection will give examples.

  \medskip 
  
 (1): {\em Dissolution remains unproven}:--- We claim, in short, that the arguments claiming to dissolve the hole argument fail. These arguments have of course many details and subtleties---as do the rebuttals of those arguments. But space is short, and most of these details and subtleties are not directly needed for this paper’s positive project of expounding the ideas of counterparts, and threading of points, needed in Part II. So we will confine ourselves to stating some  central points that have a clear connection with other Sections of this paper. So our aim here is not---cannot be---to convince all readers that the arguments fail; but just to say where we stand, and also to give newcomers a glimpse of the issues in this literature. 

First, in (1A), we will articulate one main line of argument in \cite{Weatherall_hole}, and then endorse the reply by \cite{Pooley_Read}. Then in (1B) we will give three further examples from the literature of this line of reply. Finally in (1C), we suggest a reason why debate about the hole argument continues: a reason that will lead in to our (2).

\medskip

(1A): Some philosophical jargon (which is widely used) will be helpful. The ways that the two models in the hole argument differ, namely over which point in the hole $H \subset M$ has which qualitative profile, are called {\em haecceitistic differences}.\footnote{\label{explainhaeccm}{The adjective `haecceitistic’ derives from {\em haecceitism}. It is the doctrine that some (maybe even all) objects each have a  property that no other object has, even though each of these objects may match some other object in all {\em qualitative} respects. Such non-qualitative properties, each unique to their instance, are called {\em haecceities}, and are naturally expressed with proper names that encode no qualitative information, like ‘is Humphrey’ (cf. the example in footnote \ref{Hubert}), or in logical notation `= $a$’. (Beware: philosophers’ usage of `property’ varies. Some require a property to encode some (though perhaps logically weak) qualitative information. But such philosophers can still accept haecceitistic differences, by allowing that two objects can be distinct {\em solo numero}, i.e. without any property instantiated by one but not the other.) }} 

As we have said, Weatherall takes it to be standard mathematical practice to disregard such differences, since isomorphism is to be the method of comparison between models. This goes along with his rejecting the identity map on the manifold $M$ as a method of comparison; and with our taking him to be an advocate of the drag-along response. Thus in one of the main passages expressing this view, he writes: `mathematical models of a physical theory are only defined up to isomorphism … One consequence of this view is that isomorphic mathematical models in physics should be taken to have the same representational capacities. By this I mean that if a particular mathematical model may be used to represent a given physical situation, then any isomorphic model may be used to represent that situation equally well’ \cite[p. 332; cf. also p. 338]{Weatherall_hole}. 

 As we said in this Section’s preamble, Weatherall’s paper led to a debate that continues: which we of course cannot resolve, and we can only give our view of it. So far as we know, the first article to take a similar position to Weatherall was \cite{Fletcher_hole}. Thus the last assertion of Weatherall, just quoted, gets a well-nigh synonymous formulation (and endorsement) by Fletcher, who writes: `if two models of a physical theory are mathematically equivalent, then they have the same representational capacities’ \cite[p. 233]{Fletcher_hole}. For this, Fletcher suggests a handy acronym, `REME’; (for `representational equivalence by mathematical equivalence').

To this, we endorse the reply by \cite{Pooley_Read}, who write (using our notation):
\begin{quote} 
[though] REME is a legitimate principle of mathematical representation .. . [it] fails, by itself, to block the hole argument, or to rule out the allegedly problematic use of isomorphic models deployed in standard presentations of the argument [i.e. the use of the identity map on the manifold $M$ as a method of comparison]. All that this use requires is that if model ${\bf M}_1$ (say) is taken to represent a particular possible world $W$, then ${\bf M}_2$ [i.e. the model produced by drag-along, as in Section \ref{ssecEin13}] represents (or, better, may be taken to represent, perhaps by further specifying the representational context) a distinct (but merely haecceitistically distinct) possibility $W’$. This is all completely compatible with ${\bf M}_1$ and ${\bf M}_2$’s having identical representational capacities. That merely requires that ${\bf M}_2$ might equally well have been chosen initially to represent $W$ (in which case, some other hole-diffeomorphic model, related to ${\bf M}_2$ just as ${\bf M}_2$ is related to ${\bf M}_1$, could be taken to represent $W’$). \cite[Section 5]{Pooley_Read}
\end{quote}
In short, the reply is that being precise about one’s assumptions about models’ representational capacities (though of course laudable) is not enough to rule out someone, viz. the advocate of the hole argument, taking the identity of the points to have representational significance. To put it in this paper’s jargon: precision about representational capacities does not show the drag-along response is compulsory, or that ({\em Distinct}) is untenable; (recall footnote \ref{Alice}). 

\medskip

(1B): This line of reply (and the broader moral that philosophy cannot be read off mathematics) recurs in the literature. 
Here are some examples: earlier ones by Pooley and others, and a current one by Menon and Read.

In an earlier paper, Pooley makes essentially the same reply, and connects it with the topic of determinism: saying as we did in (3) at the end of Section \ref {sssec:cpart} that a philosopher who believes that possibilities (i.e. in philosophers’ jargon: possible worlds) can differ merely haecceitistically can accept that general relativity is indeterministic. Thus he writes:
\begin{quote}
In effect, both authors [i.e. Weatherall and Fletcher] argue that, if there are pluralities of merely haecceitistically distinct possibilities, the mathematical formalism of GR, correctly interpreted, is necessarily indifferent to differences between them. But this just means that GR does not distinguish between any two elements of such a plurality; both will count as physically possible according to GR or neither will. And that, of course, is just to admit that, according to any metaphysical view committed to such pluralities, GR is indeterministic. The indeterminism cannot be avoided by remaining loftily above the metaphysical affray. \cite[end of Section 4]{Pooley_routledge}
\end{quote}

Other earlier papers making the same or similar replies include: (a) Roberts in \cite{Roberts_disregarding}, and \cite[especially pp. 255 and 261]{Roberts_hole}; Roberts argues that his condition {\em weak Leibniz equivalence}, which is well-nigh synonymous with Fletcher’s REME, is irrelevant to, i.e. powerless to block, the hole argument; and (b) \cite[Sections 3.3.2 against Weatherall, and 4.1 against Fletcher]{Arledge2019}.

Recently, the same line of reply occurs in \cite{ReadMenon_hole}'s critique of \cite {HalvorsonManchak_hole}'s attempt to support \cite{Weatherall_hole}. Most of their critique is devoted to showing (we think rightly) that Halvorson and Manchak’s efforts are misdirected.\footnote{\label{TushJamesVsHM}
For Halvorson and Manchak's   idea is that Weatherall's dissolution of the hole argument has a loophole: a loophole that they claim to close by invoking a theorem that isometries of Lorentzian manifolds are ``rigid” in the same manner as analytic complex functions, i.e. such an isometry is determined by its action on an open set. Their idea is that Weatherall’s requirement that points in distinct models can only be identified by invoking an isometry will run in to trouble if there is more than one such isometry; (cf. the problem with symmetric models we rehearsed in (3) at the end of Section \ref{sssec:dar}). So they cite such a theorem (viz. Theorem A1 of \cite[Appendix]{Geroch_limits}, and claim to complete Weatherall’ analysis and so ``close” the hole argument. 

Most of \cite{ReadMenon_hole}’s effort goes into showing that this concern about uniqueness of isometries (and so Geroch’s rigidity theorem) is irrelevant to the hole argument. The issue turns on our comment in Section \ref{ssecEin13} (after stating general covariance): that one must distinguish two senses of the word `isometry’ (or more generally, of `isomorphism’). (Cf. Menon and Read’s `isometry$_1$’ and `isometry$_2$’ in \cite[Section 2.2]{ReadMenon_hole}.) In short: a model $(M,g,T)$ can be ``wrinkly” i.e. have no automorphisms; yet, by construction, it is isomorphic to any model built by dragging-along $g$ and $T$ by a diffeomorphism $d$ of $M$.} But once this misadventure by Halvorson and Manchak is set aside,  \cite{ReadMenon_hole} address (in their Section 3.2, `Reopening the hole argument’) the questions that remain: are Halvorson and Manchak’s (and Weatherall’s) tenets justified? And do these tenets (whether or not justified) block the hole argument? They argue, using the above line of reply, that for both questions, the answer is `No’.

\medskip

(1C): Agreed, these replies are not definitive. Controversy continues: the jury is still out. For example, Fletcher addresses the objection that REME does not block the hole argument \cite[p. 239, start of Section 5.2]{Fletcher_hole}, but believes he can surmount it. Again, we think he cannot, for reasons that are spelt out by both (a) \cite{Roberts_hole} in his rebuttal of what he calls {\em strong Leibniz equivalence}, and (b)  \cite{Pooley_Read} in the remainder of their closing Section 5, which also discusses Roberts---and urges the relevance of appealing to counterparts, as we advocated in Section \ref {sssec:cpart}.\footnote{\label{SamVsGordon}{ We should note that Fletcher also makes an important distinction between REME and the logically stronger principle that `if two models of a physical theory are mathematically equivalent, then there is a unique physical state of affairs that they represent equally well’ \cite[p. 231]{Fletcher_hole}: which he gives the acronym, `RUME’ (for `representational uniqueness by mathematical equivalence'). RUME is close to Roberts’ strong Leibniz equivalence; and we concur with  Pooley and Read’s assessment of it (in their Section 5). But here, we just note that Fletcher invokes his distinction to argue (pp. 241-246) that Belot’s point (cf. footnote \ref{Belot}) that some isomorphisms generate new possibilities, since they change asymptotic structure, is not a problem for him (nor for Weatherall). For these examples refute RUME, not REME: indeed, they do so in much the same way as some much simpler examples about a particle whose worldline can swerve (which can accelerate) at different spacetime points.}}

But we will not further try to sum up the {\em status quo} in the debate. Instead, we end by reporting (uncontentiously!) a point about Weatherall's appeal to mathematical practice. (This will also lead to (2).) Namely: in the details of this debate, it turns out that Weatherall and his kindred spirits, who seek a formal or mathematical response to the hole argument, have a different view of what is the {\em content} (even apart from the interpretation) of general relativity, than do authors such as Pooley, Read and ourselves. These differences even involve disagreement about the sorts of logical or metalogical apparatus that a formulation of general relativity could, or should, be expressed with. For details of this disagreement, cf. \cite{ BradleyWeatherall_hole}; and (in reply) \cite[Sections 5, 6]{Cudek_counter1} and \cite{Cudek_counter2}. So these disagreements lie, in effect, in the philosophy of logic: which is an academic field where physicists, and even philosophers of physics, are bound to be cautious about taking sides. But for our purposes here, it is fortunately enough to just note that once one agrees it is a contentious matter of judgment what is the content of general relativity, then it is unsurprising that debate about how to respond to the hole argument continues. 

  \medskip

(2): {\em   Mathematics is indifferent and flexible in its treatment of identity}:--  We agree that modern mathematics in all its branches focusses, at least for the most part, on the ``structures’’ or ``patterns” in its subject-matter; and this means it is usually  {\em indifferent} to the identity of the objects that instantiate the structure or pattern. Here, `indifference to identity' means, of course, indifference to the objects' intrinsic natures, and to any attributes (intrinsic or extrinsic) by means of which one might  qualitatively distinguish which object is which.   So it does not mean indifference as to whether the objects are distinct: which would undermine there being a definite number of them.

 We also agree that this indifference makes it often natural to treat the objects that correspond by an isomorphism (as argument and value of the bijection giving the isomorphism) as corresponding, or ``the same”, as urged in the quote in footnote \ref{Hartry}. (This indifference to identity, and treating objects as the same, is of course a theme in both category theory, and structuralist philosophies of mathematics;  and also in the history of pure mathematics, and especially Hilbert's influence on it.)  But though this is often so, it is not {\em always} so. Indifference makes for flexibility about identity, not for isomorphisms being the mandatory standard for it.

Indeed, there are two points here. The first is that we should beware of a bad argument for isomorphisms being the mandatory standard. The second point will be more positive. 

The bad argument targets our admitting, in Section \ref{ssec:revival}, that it is `undoubtedly plausible' that substantivalism implies ({\em Distinct}): 
since for any two distinct objects, $a$ and $b$, and any two distinct attributes such objects can have, $F$ and $G$, surely $a$ having $F$ and $b$ having $G$ is a different possibility than $a$ having $G$ and $b$ having $F$; (cf. also footnote \ref{Alice}). We similarly remarked at the start of Section \ref{ssec:3replies} that Sophistication (viz. that two isomorphic models are, or represent, the same possibility) needs fancy footwork. 

It is tempting to reply to this that, not only {\em can} Sophistication claim that {\em its} version of substantivalism denies that there are facts about which spacetime point has which qualitative profile of field-values (i.e. that its silence about “which is which” still counts as belief in the points as objects): also, it {\em must} claim this---thanks to the following argument.
\begin{quote}
Isomorphisms are {\em defined} as bijections that are required (only) to preserve ``structure”, and so they can ignore the identity of the objects they map. So Sophistication must deny that which object is which can be part of what each of a pair of isomorphic models represent.
\end{quote}
But this is a bad argument. Its error lies in its assuming that the structure that an isomorphism must preserve cannot encode the identity of the arguments of the bijection. But it can do so. For in mathematics, isomorphism and structure are both entirely formal notions. They are therefore {\em flexible}. So if we wish, we {\em can} specify that the bijection preserve for each argument a property that only it has, so that the corresponding value of the bijection (in the codomain) has it.\footnote{Recalling the jargon in footnote \ref{explainhaeccm}, we can put the point as follows: by enriching the notion of isomorphism to include preserving haecceities, Sophistication can be reconciled with haecceitism.}  

 We can put the same point  in the language of formal semantics, or model theory. Agreed: an isomorphism is a bijection from the base-set of the domain model to the base-set of the codomain model that is only required to preserve the announced ``structure”, as encoded in predicates etc. of the language that the models interpret. And agreed: this announced structure {\em usually} does not incorporate anything about the identity of the base-sets' elements. The base-sets can be disjoint, or distinct but overlapping, or the same: and if they are overlapping or the same, the bijection $f$ can either (i) map an object $o$ that is in common  to itself, or (ii) instead not do so---provided only that $o$ and $f(o)$ share their announced ``structure”, i.e. patterns of instantiation of the predicates etc. in the language.  But we {\em are} at liberty to require any features, including about the identity of base-set elements, to be part of what is to be preserved by the bijection. One just needs to introduce into the language appropriate vocabulary, e.g. predicates, to encode these features. 
 	
 	Besides, even if the features to be preserved are assumed to be the usual ones of Lorentzian geometry, we do not need to appeal to any bijection between spacetimes in order to assess whether they instantiate those  features. Each diffeomorphism-invariant quantity has the same values in two given models of spacetime iff the models are isometric; or put differently, any qualitative attribute is instantiated in one represented spacetime iff it is instantiated in the other. This lets us theoretically assess the physical equivalence of the represented spacetimes without appeal to the isometry that relates them (or even to the identity map). In this sense, Sophistication  is indifferent to how points in distinct models correspond, or are identified with each other.

Our second  more positive point is that in some contexts, mathematics {\em needs} to treat objects’ identity other than by (their corresponding by an) isomorphism.  And these contexts include the treatment of points in differential geometry, both pure and applied. 

So to sum up: indifference to identity makes for a flexible, even opportunistic, treatment of identity: it does not make for the mandatory strait-jacket of a uniform treatment. Besides, this moral applies to the case of interest to us: points in spacetime theories like general relativity.

\subsection{Examples of threading}\label{subsec:32thread}

 In this Subsection, we will give four  examples of differential geometry ``identifying" points other than by an isomorphism; (though in each example, the two points will be related by a  diffeomorphism,  as argument and  value).  We will talk of {\em threading} the points, or {\em a threading scheme}, so as to signal that there is a variety of ways to thus thread points: a variety that is not conveyed by saying  ``identifying points’’.
   Note that there is no uniform meaning to this  threading of points---and no need for such a meaning.  What matters is that  differential geometry in some contexts invokes a scheme of identification {\em other than  drag-along}.

Note that the third and fourth examples will introduce the idea of threading two points that are in two {\em non}-isomorphic spacetimes: a topic we develop in Part II.

\subsubsection{The definition of the Lie derivative}\label{sssec:Lie}
To define the Lie derivative of a field, one  drags the field along the integral curves of the vector field with respect to which one is taking the Lie derivative. This means  assuming an ``identity'' or ``correspondence"---we say: threading---of the points, independent of their field-values: so that sliding the field along a curve through the points makes sense. For if points' identity was determined by field-values (as in Section \ref{sssec:dar}'s drag-along response), the field would \textit{not}  be dragged with respect to the points. More precisely: for $d_t$ the flow of a vector field $X^a$, we would obtain for e.g. a metric $g$:\footnote{\label{Arnold} Landsman makes  essentially this comment. Like us, he opposes claims (Section \ref{subsec:31notdissolve}) that isomorphisms being a mandatory standard for identifying points vitiates the hole argument. He writes: `Much of differential geometry would be thrown away if one is not allowed to compute the pullback of a metric under a diffeomorphism; even the very definition of an isometry rests on the ability to say whether or not $d^*g(p)$ equals $g(p)$, at each $p \in M$. The usual definition of the action of a diffeomorphism on a tensor (field) would be a ``category mistake”, and with it, the Lie derivative …[notation changed]'  \cite[p. 4]{Landsman2022}.  Cf. also \cite[Sec. 2.4]{Gomes_PhD}.

Amusingly, the {\em maestro}  \cite[p. 198]{arnold1989} says that the Lie derivative is sometimes called `the fisherman's derivative', because one can think of the integral curves of the vector field as the flow lines of a fluid, e.g. a river: ``the flow carries all possible differential geometric objects past the fisherman, and the fisherman sits there and differentiates them'. Thus the stationary earth of the river-bed is the analogue of the ``fixed''---we say: threaded---points over which the fields (Arnold's `differential geometric objects') slide.} 
\be\label{eq:Liederiv}
{\mathcal{L}}_{{X}}{g}(x)=\lim_{t\rightarrow0}\frac{1}{t}({g}(x)-d_{t *}{g}(d_t^{-1}(x))\equiv 0 \; !\ee

But to do differential geometry, pure or applied, one must of course accommodate non-vanishing Lie derivatives. An obvious example is immediate from the definition in relativity theory of a rigid continuous body, as a body for which the 4-velocity field  $X$ of its material particles Lie-drags the metric: ${\mathcal{L}}_{{X}}{g} = 0$. Therefore, to describe the motion of a {\em non}-rigid continuous body, we need:  ${\mathcal{L}}_{{X}}{g} \neq 0$.  But of course, there are many more sophisticated examples: one is in the next Subsection. 

Agreed, one might object to what we have just said: that there is an alternative definition of the Lie derivative as an algebraic derivation, i.e. as a set of rules about how to apply a certain operator to various quantities defined on the manifold---a definition which does not invoke pull-backs by infinitesimal diffeomorphisms. But we submit that even with such a definition, the problem of Lie derivatives trivially vanishing would remain: at least if one keeps the standard understanding of variational symmetries in a classical field theory as induced by diffeomorphisms. For as we will see in the next Subsection's example, combining this understanding with the drag-along response implies that any Lie derivative that would emerge from the action of  a symmetry must trivially vanish.

\subsubsection{Noether’s second theorem in general relativity}\label{sssec:Noeth}

Noether's second theorem in general relativity gives a central example of our theories needing a non-zero Lie derivative. We suppose, to begin with,  that the action takes the form
\be
\label{eq:GR}
\int \d^4x \,(\mathcal{L}_g+\mathcal{L}_m) \; ; 
\ee
where we assume minimal coupling, in the sense that all contributions of the matter fields to the Lagrangian are confined to the second component, $\mathcal{L}_m$, of the total Lagrangian density. We also assume that the energy-momentum tensor for the matter fields is given by (introducing indices)
\be \label{eq:eom_matter}
T^{ab}=\frac{\delta \mathcal{L}_m}{\delta {g_{ab}}},
\ee
and that the action is given by the Einstein-Hilbert Lagrangian, $\mathcal{L}_g=R\sqrt {g}$. 

Now we would like to apply a variation along an infinitesimal symmetry,  that is along a vector field $X^a$ that is the infinitesimal generator of a flow diffeomorphism. Assuming that general flow diffeomorphisms generate symmetries of the theory,   and that $X$ has compact support so that the boundary terms vanich, we obtain:
\be\label{eq:inf_sym_gr} 
0=\delta S=\int \d^4 x((\frac{\delta \mathcal{L}_g}{\delta {g_{ab}}} + \frac{\delta \mathcal{L}_m}{\delta {g_{ab}}})\mathcal{L}_{{X}}{g}) \equiv \int \d^4 x((G^{ab}+T^{ab})\mathcal{L}_{{X}}{g_{ab}}).
\ee

With the drag-along response, i.e. with the fields {\em not} being dragged over the spacetime points, $\mathcal{L}_{{X}}{g}\equiv 0$. So \eqref{eq:inf_sym_gr} would be satisfied identically, and yield no further conditions on the quantities involved.  

But with threading, i.e. the fields being dragged over the ``fixed canvas" of points, we  get $\mathcal{L}_{X}{g_{ab}}=\nabla_{(a}X_{b)}$ (with $\nabla$ the Levi-Civita covariant derivative). Then, from \eqref{eq:inf_sym_gr}, we obtain:
\be 
\nabla_a G^{ab}=\nabla_a T^{ab},
\ee
which is satisfied even off-shell: that is, even independently of the Einstein equations. Therefore, using the Bianchi identity $\nabla_aG^{ab}=0$, we also get the off-shell relation: $\nabla_aT^{ab}=0$, i.e. the local conservation law for the energy-momentum tensor.\footnote{We stress that this argument does not show---and we do  not claim---that the drag-along response implies that energy is not locally conserved. But the argument shows, and we claim, that the drag-along response implies that a  proof of conservation cannot go via Noether's second theorem.}

\subsubsection{Limits of spacetimes}\label{sssec:limits}

 In the two preceding Subsections, we have threaded points between two isomorphic spacetimes ---though, of course, not by an isomorphism. In this Subsection and the next, we turn to threading points between two non-isomorphic spacetimes: in particular, setting aside matter fields, between non-isometric Lorentzian manifolds. 
 
 Our first example is the use of threading in the rigorous construction of the limit of a (one real parameter) family of Lorentzian manifolds, by Geroch \cite{Geroch_limits}: whose exposition we will follow.\footnote{\label{Malament}  Though we confine ourselves to Geroch’s exposition, we stress that the construction has had many applications. The obvious ones are in proofs of the stability of a spacetime: for in such a proof one needs to consider a sequence of spacetimes converging to the given one (and show that their geometries converge to its). See for example \cite[Sec. 7.6]{HawkingEllis}, \cite[p. 968]{Belot50}    and (stressing topological issues)  \cite[Section 4]{FletcherBJPS2016},  and \cite[Section 5.2]{FletcherWarsaw}.}

We note also that both this example and the next Subsection's use only relativistic spacetimes. But there are theorems stating that a non-relativistic spacetime is a limit of a suitable family of relativistic spacetimes. The best-known example is a Newton-Cartan spacetime being such a limit ``as $c \rightarrow \infty$":  the null cones get wider until they finally become degenerate, i.e. tangent to a spacelike hypersurface of the Newton-Cartan spacetime. (Cf. e.g. \cite{Malament1986}: (Proposition 1, p. 194,  and Proposition 2, p. 197, which treats matter); \cite{FletcherSHPMP2019} for an analysis stressing topological issues.) Space forbids our going into details: but it is intuitively clear that such theorems can be thought of in terms of threading. For example, consider the idea that as the null cones widen, curves through a cone's vertex that were (in earlier members of the sequence of spacetimes, with narrower cones) spacelike enter the cone i.e. become timelike. This formulation invokes a threading across the models of a sequence of points. And as in our other examples, the sheer mathematics is happy to treat this threading as identity, and so to use a single manifold for all the models. To use the obvious metaphor: we thread the points by their occupying the same place on ``the sheet of paper" in common between the models, not by their matching as regards being spacelike or timelike to the cone's vertex (a chosen origin), nor by any other matching of metrical attributes.

Geroch's main idea is that the limit of a (1-parameter) family of spacetimes  (in general, {\em not} isometric to each other), should be defined as the boundary of a certain 5-dimensional manifold $\mathcal{M}$, of which the given spacetimes are 4-dimensional leaves. For simplicity, Geroch sets aside matter fields, so that spacetimes are Lorentzian manifolds $(M,g)$. He writes the family as $(M_{\lambda}, g_{\lambda})$ with $\lambda >0$; and the 5-dimensional manifold of which the $(M_{\lambda}, g_{\lambda})$ are leaves, as $\mathcal{M}$. So the limit spacetime is $\partial  {\mathcal{M}}$ and corresponds to $\lambda = 0$. (So in this Subsection we will follow Geroch's notation, and write a spacetime as $(M_{\lambda}, g_{\lambda})$, or for short, as $M_{\lambda}$:  we do not use our previous bold-font notation ${\bf M}$.)

Geroch's motivation for seeking a definition is that `one cannot speak simply of the limit of [e.g.] the Schwarzschild solution as the mass $\lambda \rightarrow 0$’, for [as he shows] the spacetime one obtains in the limit [e.g. Minkowski or Kasner] depends on the choices of coordinates’ (p.182). He also declares at the outset (p. 181) that to give a coordinate-free definition, he will consider (i) a 1-parameter family  $M_{\lambda}$ (${\lambda} > 0$), rather than (ii) a 1-parameter family of metrics on a given manifold: for he does not wish to presuppose a scheme that identifies a point $p_{\lambda} \in M_{\lambda}$ with a point $p_{\lambda '} \in M_{\lambda '}$ (${\lambda} \neq {\lambda}'$). (We note that even (ii) could spell trouble for Section \ref{sssec:dar}'s drag-along response, since the different $M_{\lambda}$ will not be isometric.) 

He also stresses (p. 182) that `limit spacetime' will be defined without  presupposing any preferred scheme for identifying (in our jargon: threading) points between different spacetimes in the family. And he points out  (p. 182, footnote) that the natural way to define a scheme is by a congruence of curves, through the 5-dimensional manifold $\mathcal{M}$, such that each curve cuts each leaf, i.e. each 4-dimensional spacetime, just once. In our jargon, this of course means that two points in (the image of) one such curve are threaded together. So agreed: as regards Geroch's general definition of limit spacetimes, threading is {\em not} needed.  And `all the better', one might say, since the spacetimes are not isometric;  (indeed, an advocate of Section \ref{sssec:dar}'s drag-along response presumably would say this).\footnote{\label{GerochDefn}  For completeness, here is his definition. He assumes the $M_{\lambda}$ `may be put together to make a smooth (Hausdorff) 5-dimensional manifold'. Then:  `we define a {\em limit space} of $\mathcal{M}$
as a 5-manifold ${\mathcal{M}}'$ with boundary $\partial {\mathcal{M}}'$, equipped with a tensor field $g^{' \alpha\beta}$, a scalar field ${\lambda'}$, and a smooth, one-to-one mapping $\Psi$ of  $\mathcal{M}$ onto the interior of ${\mathcal{M}}'$ such that three conditions are satisfied:
 1. $\Psi$ is an isometry, i.e. $\Psi$ takes $g^{ \alpha\beta}$ into $g^{' \alpha\beta}$, and ${\lambda}$ into ${\lambda'}$;
2. $\partial {\mathcal{M}}'$ is the region given by ${\lambda'} = 0$: we require, furthermore, that   $\partial {\mathcal{M}}'$ be connected Hausdorff and non-empty;
 3.  $g^{' \alpha\beta}$ has signature $(0, =, -, -, -)$ on $\partial {\mathcal{M}}'$.' (p. 182)} 

But threading comes into its own (and the limited scope of the drag-along response is shown) in the {\em application} of the definition. For as Geroch says, the general definition is not `useful for actually writing down limits' (p. 183).  To write down limits, one needs to use a family of frames (tetrads) on the $M_{\lambda}$, smoothly varying along the curves of a congruence of curves  in $\mathcal{M}$, each curve cutting each leaf once. Thus each such congruence specifies a threading relation: $p_{\lambda} \in M_{\lambda}$ and $p_{\lambda '} \in M_{\lambda '}$ are threaded together iff they lie on the same curve of the congruence. (Of course there are many---continuously many---congruences, so the threading relation is fiducial.) Thus Geroch writes:
\begin{quote}
By a {\em family of frames} in  $\mathcal{M}$ we mean an orthonormal tetrad $w(\lambda)$ of vectors tangent to $M_{\lambda}$ and attached to a single point $p_{\lambda} \in M_{\lambda}$, for each $\lambda > 0$, such that the $w(\lambda)$ vary smoothly along the smooth curve in $\mathcal{M}$  defined by the points $p_{\lambda}$. [In our jargon, this curve of the congruence is one set of points threaded together.] 

Let  ${\mathcal{M}}'$ be a limit space of $\mathcal{M}$, and let $w(\lambda)$  be a family of frames which assumes a limit---i.e., approaches a frame $w(0)$ at some point $p_0 \in \partial {\mathcal{M}}'$---as $\lambda \rightarrow 0$. Let us represent points in $M_{\lambda}$ in a neighborhood of $p_{\lambda}$ in terms of the system of normal coordinates based on $w(\lambda)$. In terms of these coordinates, the components of the metric tensor in the $M_{\lambda}$  approach a limit as $\lambda \rightarrow 0$, and the limiting components are precisely the components of $g^{ab}(0)$ in $\partial {\mathcal{M}}'$ in a neighborhood of $p_0$. Thus, the family of frames $w(\lambda)$ uniquely defines the limit space ${\mathcal{M}}'$, at least in a suffiiciently small neighborhood of $p_0$. (p. 183)
\end{quote}
To sum up: in order to explicitly construct limits of spacetimes, one needs to thread points between non-isometric spacetimes.\footnote{\label{Curiel}{We are not the first to invoke this construction by Geroch in connection with the hole argument. In a paper written at about the same time as \cite{Weatherall_hole}, \cite{Curiel2018} does so. About the hole argument, he is sympathetic to Weatherall (whom he cites), saying: `Thus the hole argument is obviated by the fact that the application of the diffeomorphisms to the manifold cum metric results only in a different presentation of the same intrinsic physical structure, and so the worry about indeterminism evaporates, doing away with the dilemma' (p. 453). But he also argues more generally that for the question of whether spacetime points exist (or exist independent of their metric relations) to be fruitful, we must place it in a specific scientific context: and limits of spacetimes {\em a la} Geroch provide just such a context. Hence Curiel gives criteria for a Yes answer to the question in terms of such limits (his Definitions 1 and 2, p. 463).} }

\subsubsection{Quantum reference frames}\label{sssec:qrf}

 Our last example opens that Pandora’s box, the interpretation of quantum theory: especially in relation to superpositions of spacetime geometries, induced by superpositions of matter distributions. Of course, we must be brief, and cannot hope to settle any controversial matters. But our main point—that this literature invokes the idea of threading: indeed, of rival threadings---holds good, irrespective of these controversies. 

So we will first present the general issue; and report a proposal of Penrose. Then we report a recent critique of his proposal by Giacomini and Brukner, that uses the framework of {\em quantum reference frames}. But again: what matters for us is agreed by the disputants: the need for threading points between non-isometric spacetimes.

The first point to make is the obvious one. In combining quantum theory with any spacetime theory (such as general relativity) in which spacetime is dynamical, i.e. responsive to matter, the question arises whether the metric (or more generally: geometrical or gravitational) degrees of freedom are to be given a quantum treatment.  We shall (like Penrose, Giacomini and Brukner) aim for a    treatment in which superpositions of distributions of the masses sourcing the gravitational field imply some sort of superposition of non-isometric spacetimes. There are then two questions. First, and fundamentally: how should we understand such a superposition? Secondly, what should we say about cases where two (or more) mass distributions that are components of the superposition are macroscopically distinguishable, so that the ensuing difference in the two geometries is also macroscopically distinguishable?  

Over the years, Penrose has  proposed answers to both questions (as of course,  other authors have):
answers that vividly raise the issue of ``identifying", i.e. threading, points between non-isometric spacetimes.  For example, Penrose writes: `The basic principles of general relativity—as encompassed in the term ‘the
principle of general covariance’ (and also ‘principle of equivalence’)—tell us
that there is no natural way to identify the points of one space-time with
corresponding spacetime points of another' \cite[p. 591]{Penrose_gravi_collapse}. 

Now, each (classical) metric (or gravitational) field involved in the superposition will have a different timelike Killing
vector, and thus a different direction of preferred time-translation. Penrose then argues that because of
the relation between timelike Killing vectors and energy conservation laws, the fact that the former are not
well-defined means that the global state is not stationary; and the difference in the energies implies an energy-time 
uncertainty principle, which makes the superposed quantum state unstable. This is the basic idea of Penrose's  gravitationally-induced state-reduction.
   
Recently, the framework of {\em quantum reference frames} (see \citep{Brukner_super, Giacomini2020, Hardy2020} and references therein) has been applied to these proposals; with \cite{Brukner_super}, in particular, criticising the argument for gravitationally-induced ``collapse of the wave packet”. They point out that there is no single reference frame transformation which makes the spacetime locally Minkowskian along a geodesic in such a superposition.  This, according to them, is the source of the  apparent violation of the  weak Equivalence Principle, i.e., the violation of the universality of free-fall.

 In more detail: they say that Penrose’s argument rests on the (implicit) assumption that ``all coordinate systems are in a classical relation to each other. With this assumption, it is impossible to find a classical, local coordinate system in which the metric can be made locally Minkowskian for both configurations of the mass'' (ibid. p. 4). They then propose, using their framework,  that one should anchor the labelling of  spacetime points onto the trajectories of the masses involved. This they take to disarm Penrose's argument: the Equivalence Principle holds  in each component and hence in the entire superposition.

For our concerns, what matters is of course the ``anchoring of the labelling''. Thus for the case of a single mass that acts as a probe particle for the metric/gravitational field (the particle being, of course, present in all elements of the superposition), they take 
the different spacetime coordinates of the particle in different manifolds to all correspond to the same physical point,  as
specified by the location of the particle (ibid. p. 5). 
In our language, this  amounts to using the values of e.g. a mass-density function for the particle (thought of as the ``same in the different possible worlds"!)  so as to define how to thread points. 

In Part II, we shall return to this suggestion, using our counterpart-theory framework for treating non-isometric spacetimes. For now, we close by just noting that  \cite{Brukner_super} implement the idea that ``threading is defined by mass-density" by having some object---other than the probe particle---define a quantum reference frame, by its being placed at the origin of a coordinate system. Thus they write:     
\begin{quote}
Such relative coordinates correspond to the distance between any physical system and the initial quantum reference frame, defining
[...] 
the origin of a coordinate system. This allows us to operationally identify points belonging to different
spacetimes, and meaningfully write the superposition state by adopting a relational view: the quantum reference frame [...]
is always at the origin of the coordinate system, and the quantum states of [the probed particles]
are their relative [...] 
states, in a position basis. [...] In this view, there is no contradiction in having
different Killing vectors which are in a superposition in each amplitude. \cite[p. 4]{Brukner_super}\footnote{Incidentally,   there is a technical issue. One glimpses it in this quotation's mention of relative coordinates and relative distances (relative to the object at  the origin of  the quantum reference frame).    This suggests that Giacomini and Brukner's framework secures a unique Fermi coordinate system along the trajectory of the probe particle that is identified across the two (or more) spacetimes. But distances to a given object  are not sufficient to determine a local coordinate system: one also needs information about directions. In other words: to define a unique  Fermi coordinate system along the particle trajectory, we need more than just the metric and the trajectory---we need some kind of anisotropy. Agreed:  Brukner and Giacomini seem more interested in  existence,  not uniqueness, of Fermi coordinates in each member of the superposition. Moreover, uniqueness can be obtained by assuming cylindrical symmetry around the particle trajectory. } 

\end{quote}

 \section{Conclusion}\label{concl}

This concludes our review of the philosophical discussion of the hole argument. Since Section \ref{ssec2aims} already gave a summary of our main themes and claims, we will not do so again here. But it is worth stressing three points. For they concern the central theme, whether isomorphic models must be regarded as representing the same possibility (called `Sophistication’); and they all look ahead to Part II.

(1): {\em The prospects for Sophistication}:---  We have emphasised that although we reject the drag-along response (Sections \ref{sssec:dar} and \ref{sec:3dissolve?}), we are sympathetic to Sophistication; (cf. (1) and (2) at the end of Section \ref{ssec2aims}, and the preamble to Section \ref{ssec:3replies}). But we do not claim to add to the {\em philosophical} case in favour of Sophistication: neither in this paper, nor in Part II. In particular, the main merit of Part II’s fibre bundle framework, as regards Sophistication, will be to place discussion of it in a wider context.

(2): {\em Counterparts and threading …}:---  This paper has set the stage for Part II’s endeavour, viz. comparing non-isomorphic spacetimes. In particular, the ideas of counterparts (Section \ref{sssec:cpart}) and of threading points (Section \ref{subsec:32thread}) will be important tools for such comparisons.       

(3): {\em … in a fibre bundle of spacetimes}:---  We will see that these two ideas can be implemented naturally in the fibre bundle of spacetimes. For example, a gauge-fixing condition in the fibre bundle, i.e. a choice of a section of the bundle that cuts across the fibres, can be  specified by intuitive counterpart relations between spacetime points.  (We say `intuitive', because these relations can invoke long-established ideas that points correspond to one another by having suitably matching, or nearly matching, values of physical quantities.)  Besides, the technical setting enables us to define counterparthood, i.e. degrees of similarity, for entire spacetimes. And we can then show that counterparthood for spacetimes has a natural transformation property;  and even that this transformation property reveals a grain of truth in the claims (Section \ref {subsec:31notdissolve}) that isomorphism is the mandatory standard for when points in two models should be (that dreaded word!) ``identified".

\subsection*{Acknowledgements}
For comments on this material, we are grateful to: the audience, and especially the organizers, at the DICE conference in Castglioncello; the audience and organizers at the August 2022 QISS meeting;  the participants at the LSE-Cambridge Philosophy of Physics Bootcamp discussion group; to seminar audiences at Bonn, Bristol, Oxford and Warsaw;  and to Caslav Brukner, Frank Cudek, Sam Fletcher, Flaminia Giacomini, Ruward Mulder, James Read and Isaac Wilkins  for comments on a previous version.


\end{document}